# Bohr's way to defining complementarity[(*)]

Alberto De Gregorio
via del Colle di Mezzo, 21 – 00143 Rome, Italy (tel. no. +39 (0)6-5042895)
e-mail: a_degregorio@yahoo.it

**Abstract.** We go through Bohr's talk about complementary features of quantum theory at the Volta Conference in September 1927, by collating a manuscript that Bohr wrote in Como with the unpublished stenographic report of his talk. We conclude – also with the help of some unpublished letters – that Bohr gave a very concise speech in September. The formulation of his ideas became fully developed only between the fifth Solvay Conference, in Brussels in October, and early 1928. The unpublished stenographic reports of the Solvay Conference suggest that we reconsider the role that discussions with his colleagues possibly had on Bohr's final presentation of the complementary sides of atomic physics in his 1928 papers.

> *There casually happened (as was usuall) several discourses at times between these Gentlemen, the which had rather inflamed than satisfied in their wits the thirst they had to be learning; whereupon they took a discreet resolution to meet together for certain dayes, in which all other business set aside, they might betake themselves more methodically to contemplate the Wonders of God in Heaven, and in the Earth.*
>
> Galileo Galilei (*Dialogues on two World Systems: to the Judicious Reader*, tr. Th. Salusbury)

> *Besides, there is no scientific work that one man alone can write.*
>
> Bertolt Brecht (*Life of Galileo*: Scene XIV)

## 1. Introduction

As Galilei and Brecht vividly remind us, science is intrinsically relational – like any other human activity. The right enthusiasm and admiration, we may feel before any milestone achievement in physics and the person who eventually reached it, do not allow us to underestimate the more or less thick, more or less evident web of human relationships and professional interactions which allowed that result to take shape. As a matter of fact, sometimes the personality of the discoverer may instead

---

[(*)] This paper is based on, and further develops, *Bohr's talk at the 1927 Como Conference,* a talk the author gave at the *Second International Conference on the History of Quantum Physics* in Utrecht in July 2008, and on which De Gregorio & Sebastiani (2009) is also based.

overshadow the whole context from which his discovery arose.

Niels Bohr gave his first public accounts of complementarity in the second half of 1927. He had two main occasions: the Conference for the centenary of the death of Alessandro Volta, in Como in September, and the Fifth Solvay Conference, in Brussels in October.

A wide and detailed account of the genesis of the idea of complementary features of quantum physics is in the sixth volume of *Niels Bohr collected works*, edited by Jørgen Kalckar and endowed with numerous letters and handwritten documents. Sometimes Kalckar explicitly states there, and more often implicitly assumes, that already in his speech at the Volta Conference in Como did Bohr thoroughly expound on complementarity. Cassidy, Pais, and Mehra and Rechenberg, seem to conform to Kalckar's assumption as well. However, no document is provided, showing that Bohr had fully developed his idea of complementary aspects of atomic theory at the time of the Volta Conference, nor that his speech in Como was exhaustive to any extent.

In this paper, we shall review Bohr's early account of the complementary features of the description of nature. Based on documental evidence, we will move along two lines. Our first course will be that Bohr's presentation of his ideas was still at an early stage of gestation in Como. We shall collate two documents: a draft titled *Fundamental problems of the Quantum theory*, sketching out some essentials of complementarity and written in Como on September 13 by Bohr, and the stenographic report of the speech that Bohr gave at that Conference on September 16. On the basis of this comparison, and with the help of some unpublished letters – kept at the Niels Bohr Archive – between the Committee of the Conference and Bohr, we shall conclude that the latter's speech in Como in September was very concise. It contained only embryonic ideas about complementary features of quantum theory. In Brussels in October Bohr made in public an improved exposition of his views on the complementary sides of atomic physics, which he further substantiated in the detailed accounts he would publish in the Proceedings of the Volta Conference, in the Proceedings of the Solvay Conference, on *Naturwissenschaften*, and on *Nature*, in 1928.

As for our second course, the role of Bohr's interaction with his colleagues will be reviewed in the light of the meagre content of Bohr's speech at Como. We shall start from a manuscript of October 12-13, 1927, which was discussed with Darwin and Pauli by Bohr in Italy, after the Volta Conference, and which improved the unsubstantial manuscript of September 13: we shall consider that these discussions allowed a substantial improvement of Bohr's presentation of complementary sides of nature, rather than a mere refinement of an already developed presentation. Bohr further improved the formulation of his views on the eve of the Solvay Meeting, or maybe in the very days of the Meeting. In Brussels, he discussed topics he had not included in his October manuscript (for example the gamma-ray microscope); he apparently treasured also discussions with his colleagues:

some issues examined with his peers (for instance the Stern-Gerlach experiment or the scattering of alpha-particles) are to be found in his final presentation.

**2. The Volta Conference**

"Dear Colleague, next year the town of Como will celebrate the centenary of the death of […] Alessandro Volta. […] We pray you that you would honour the Congress with your presence […]. We would very much appreciate it if you would agree to communicate an original work of yours at the Conference."[1] With a letter from Milan, dated July 5, 1926, the Committee for the Celebrations in honour of Volta invited Bohr to take part in the Como Conference of September 1927. Bohr was expressly asked, more than one year in advance, to deliver a speech during the Congress.

Bohr did not answer this letter (yet he answered H.A. Lorentz's invitation, of June 7, 1926, to attend "another '*Conseil de Physique*' to be held in Brussels [...] in October 1927"[2]). On June 14, 1927, almost one year after the invitation to Como, the President of the Committee sent Bohr another letter, from Bologna: "Dear Colleague, your Academy has just communicated to us that you and Dr. Kramers will come on their behalf to the Como Conference. […] We sent an invitation letter to you last July, one of the first being sent […]. I would regret very much if you had not received that letter. […] You can find the updated list of the contributions here attached. […] They will be published in a volume, and of course can be longer than the oral speeches."[3] Obviously, Bohr had received the letter of July 5, 1926, now kept at the Bohr Archive. Even if the Committee had not received any answer from Bohr, by June 1927 he had already accepted to take part in the Conference on behalf of the Danish Academy. Obviously as well, the participants, Bohr included, were expected to deliver speeches shorter than the subsequent printed papers.

On June 20, 1927, Bohr was ready to communicate the title of his contribution: *Fundamental problems of the Quantum theory*.[4] The same title would be given to the already mentioned manuscript written in Como on September 13,[5] but that title would be changed into *The quantum postulate and the recent development of atomic theory* in the Proceedings of the Conference.[6] Bohr also asks Q. Majorana information about the suitable length of his speech and about its publication.

---

[1] Letter to Bohr, dated Milan 7/5/1926 (*Niels Bohr Archive*, Folder VOL27). The letter was signed by Quirino Majorana (President of the Committee for Scientific Conferences), Enrico Musa (General Clerk of the Executive Committee), and Enrico Médail (President of the Executive Committee). The letters between Bohr and the Committee are in French (author's translation).

[2] Bohr answered to Lorentz on 6/24/1926. See Mehra & Rechenberg (2000, p. 175).

[3] Letter of Q. Majorana to Bohr, dated Bologna 6/14/1927 (*Niels Bohr Archive*, Folder VOL27).

[4] Rough copy of Bohr's letter to Q. Majorana, dated Copenhagen 6/20/1927 (*Niels Bohr Archive*, Folder VOL27).

[5] Kalckar (1985, pp. 75-80).

[6] Bohr (1928a).

On July 4, 1927, the Committee sent Bohr another letter, providing us with further, precious information: «Dear Colleague, […] given the high number of planned speeches, it would be fine to limit them to twenty minutes each. […] We would be very grateful to you, if you sent us in advance a short summary of your talk. You would also facilitate the duties of the Committee if, before the Congress starts, you sent us the definitive version of the paper to be published. It would help make the composition faster».[7] Three pieces of information are worth noting: as early as July 4, 1927, Bohr was asked *i*) to send a short summary in advance, and *ii*) to provide a complete paper before the Congress started; *iii*) the talks were expected to be twenty minutes long each.

On August 24, Bohr answered that he regretted that he had been ill and could not finish the paper he was asked for. However, the Abstract of his talk was ready, and he attached it: «<u>Fundamental problems of the quantum theory</u>. In connection with the recent remarkable development of the quantum theory a discussion is attempted of the well-known paradoxes of atomic physics, especially in their relation to the problem of the space time coordination».[8]

We may ask ourselves a few questions now: given the tumultuous advances of quantum physics, when did Bohr exactly resolve that he would write a paper about the principles of quantum theory? What did he really say in Como? Who did Bohr interact with, before he published *The quantum postulate and the recent development of atomic theory* in 1928?

## 3. Planning a paper dealing with the general principles of quantum theory

W. Heisenberg had been lively engaged in debates about the foundations of quantum theory with Bohr[9] since 1926, when he went to Copenhagen. In his 1927 celebrated paper about uncertainty Heisenberg announced that "the most recent investigations of Bohr [...] are soon to appear in a paper on the conceptual constitution of quantum theory."[10] He confirmed with W. Pauli: "Bohr wants to write a general paper on the 'conceptual basis' of the quantum theory, from the point of view 'there exist waves and particles'."[11] It was May, 1927.

Bohr himself had already confided to Fowler on October 26, 1926: "After the discussion with Schrödinger [who had just been in Copenhagen] it is very much in my mind to complete a paper dealing with the general principles of quantum theory."[12] To Einstein, on April 13, 1927: "The content [of Heisenberg's paper] is closely related to the questions that I have had the great pleasure of discussing with you a number of times." Bohr recalls the issues of the finite extension of the waves, related to the indeterminacy in the wavelength, as well as the limited cross section of the wave train, similarly related with uncertainty in the parallelism of the rays; he also mentions the Doppler effect. Then he continues: "For a long time I have had the intention of trying to clarify my thoughts on the general question in a small article, but the development runs so tempestuously, that everything anew becomes quite commonplace. Still I hope soon to finish such an article."[13]

---

[7] Letter of Giulio Dalla Noce, on behalf of Q. Majorana, to Bohr, dated Bologna 7/4/1927 (*Niels Bohr Archive*, Folder VOL27).

[8] Rough copy of Bohr's letter to Q. Majorana, dated Copenhagen 8/24/1927 (*Niels Bohr Archive*, Folder VOL27).

[9] Cassidy (1992, pp. 264-266).

[10] Heisenberg (1927, p. 198), in Kalckar (1985, p. 20).

[11] Pais (1991, p. 309).

Kalckar states that Bohr commenced to write a note as an answer to a letter, which Campbell had published on Nature on May 28, 1927. Moreover, he recalls that "according to Klein a last effort to finish the [answer to Campbell] was made on the very eve of Bohr's departure for Como."[14]

Given all these events, one might conclude that uncertainty shrouds the exact date when Bohr resolved he should start writing, and actually wrote, the paper he had repeatedly announced. Interestingly enough, the first document, explicitly dated, where we can find reference to "complementary aspects of experience that cannot be united into a space-time picture based on the classical theories" is a sketchy manuscript of July 10, 1927. July 10... just a few days after July 4: Bohr should have just received the third letter[15] of the Committee, asking him for a brief summary!

Bohr's writing an articulated manuscript, worth being sent to the Committee, came out to be a painful enterprise: "I feel dreadfully ashamed – he wrote to C.G. Darwin on August 29 – that not yet to have finished [sic] any paper about the general views on the quantum theory about which we talked so often. [...] I hope in Como to be able to give a reasonably clear account of my views." Bohr was missing interaction with his colleague, and was looking forward to meeting him again in Como in order to "renew our discussions."[16]

Bohr's speech at the Volta Conference was planned for September 16. A manuscript dated September 13 is kept in the Archives.

**4. Simply a matter of time**

---

[12] Letter of Bohr to Fowler, of 10/26/1926, in Kalckar (1985, pp. 14-15).

[13] Letter of Bohr to Einstein, of 4/13/1927 (Kalckar 1985, pp. 21-23). Note the word "commonplace" that Bohr uses to denote his own regret.

[14] Kalckar (1985, p. 28).

[15] Mehra and Recheneberg, instead, hold that Bohr was acknowledged of the Volta Conference only *after* he had started writing the manuscript. Mehra & Rechenberg (2000, part I, p. 188n).

[16] Letter of Bohr to Darwin, of 8/29/1927. A rough copy of the letter is kept in the Niels Bohr Scientific Correspondence (*NBSC*) at the Niels Bohr Archive. A microfilmed copy of the *NBSC* is owned by the Accademia nazionale delle scienze detta dei XL in Rome (the letter to Darwin is in microfilm no. 9).

*Fundamental problems of the quantum theory* is the title of the eight page manuscript of September 13,[17] the same Bohr communicated to the Committee with the letter of June 20 and with the Abstract of August 24. Does this manuscript exhaust the whole content of Bohr's speech in Como? Most historians say it does not.

Kalckar assumes – sometimes explicitly, some other times implicitly – that Bohr's speech almost reproduced the content of the much longer paper that he would publish in the Proceedings of the Conference in 1928, *The quantum postulate and the recent development of atomic theory*.[18] It is a twenty-four pages paper, divided into five sections (Kalckar concedes that the fifth section might also be excluded from Bohr's speech in Como: eighteen pages instead of twenty-four).[19] No document is provided in support of these views. Let us now go through Kalckar's tentative argument, which runs as follows: "The manuscript [of September 13] lacks substance", *then* "it is hard to imagine that Bohr in his delivery of the lecture would have restricted himself to the indication given here."[20]

It is evident that the argument does not work. Bohr's talk could be very concise as well, or could be very extended: simply we do not know at this stage. In particular, given only that "the manuscript lacks substance," this fact alone cannot exclude that also Bohr's lecture was meagre.

Anyway, both D. Cassidy and A. Pais seem to agree with Kalckar's point of view.[21] Also J. Mehra and H. Rechenberg hold that "it must be assumed that Bohr went in his lecture beyond what he had written in the above manuscript, which thus gives only an indication of what he presented in more detail before his audience."[22] Again, they do not provide any document to support their assumption. So, why "it must"?

In order to substantiate our conclusions by documental evidence, let us begin examining our information about the Conference.

Here is the Program:

    Sunday, September 11: Inauguration Ceremony

    Monday 12: Experiments on the Structure of Matter

---

[17] Kalckar (1985, pp. 75-80).

[18] Bohr (1928a). The same title Bohr gave to the slightly extended version published on *Nature* and – translated – on *Naturwissenschaften* and on the Proceedings of the Fifth Solvay Conference in 1928. Bohr (1928b,c,d).

[19] Kalckar (1985, p. 30).

[20] Kalckar (1985, p. 29).

[21] According to Cassidy, in October, "Bohr presented the promised paper at Como." Also Pais seems not to have many doubts that the published paper actually reproduces Bohr's speech. Cassidy (1992, p. 248). Pais (1991, pp. 309-313).

[22] Mehra & Rechenberg (2000, pp. 193-194).

> Tuesday 13: Electricity and its Applications
>
> Wednesday 14: Electrology
>
> Thursday 15: Physical Optics
>
> <u>Friday 16: Theories on Matter and Radiations</u>
>
> Saturday 17: Overview of ongoing works (in Pavia)
>
> Monday 19: Solemn Commemoration of Volta and Conclusion Ceremony (in Rome) [23]

Note that the only day expressly devoted to topics related to quantum theory was September 16. The other days, apart from summaries and commemorative speeches, were devoted to experimental issues, to electricity and its applications, to electrology, and to optics. We should not think of the Volta Conference as of such a thematic Congress as for example a Solvay Meeting.

M. Born, A. Sommerfeld, T. Levi-Civita, P. Debye, M. von Laue, Eddington, H.A. Kramers (besides P. Straneo and G. Gianfranceschi) gave their talks the same day as Bohr, for a total of ten lecturers on a single day. In addition, discussions by Lorentz, E. Fermi, E.H. Hall, M. Planck, O.W. Richardson, O.M. Corbino, J. Frenkel, Heisenberg, Pauli, M. de Broglie, and again Sommerfeld, Born and Kramers, were recorded on that same September 16. Also given the rank of the lecturers, not to say of the persons taking part in the discussions that Friday, it is straightforward that all speakers had preliminarily been asked to limit their talks to twenty minutes each. So had Bohr.

In order to reach more cogent conclusions, we are now going through documents relevant to the very content of Bohr's speech.

**5. Collating documents: the real substance of Bohr's speech in Como**

Given that Bohr's allotted time was only twenty minutes, what was the real extension of his talk? The stenographic report of Bohr's talk in Como would best acquaint us with the content of Bohr's speech. Fortunately indeed, the Niels Bohr Archive keeps that report. However, Kalckar writes that "the stenographic report in possession of the Archive seems incomprehensible."[24] In fact *it seems*; but *it is not* completely incomprehensible. However true it be that many blanks and altered words strongly undermine its comprehensibility, still it is the very stenographic report that will help us solve the puzzle of the content of Bohr's speech. We only need to consider the report together with the manuscript of September 13, and collate them.

The manuscript of September 13 includes eight pages. The first seven ones contain a developed though concise text. The last page consists of just a list of topics, which Bohr does not enter in any detail and will instead develop in his paper on the Proceedings of the Volta Conference. The first seven pages of Bohr's manuscript *Fundamental problems of the quantum theory* may be divided into three parts. The first one deals with the limitation of classical concepts for the study of atomic phenomena. In particular, the issue of the dependence on the tools of measurement is put forward. In the second part, Bohr goes back to the wave and particle theory of light, and to the wave theory of material particles, obtaining the relations $\Delta t \Delta E = h$ and $\Delta p \Delta l = h$. In the third part, a few simple examples are discussed.

---

[23] *Atti del Congresso internazionale dei fisici* (1928, vol. I, p. XII).

[24] Kalckar (1985, p. 29 n. 28).

A plain correspondence of this manuscript with the stenographic report of Bohr's speech of September 16 comes out. Some slight differences are of course unavoidable (and there are even cases in which such correspondence is not obvious at all), but on the whole the collation is absolutely persuasive. We are now going through just few examples (a detailed collation is reported in the Appendix).

The first part of the manuscript of 9/13 (pp. 1-2) opens with the acknowledgment of a "fundamental limitation in our classical physical ideas when applied to atomic phenomena." The interaction of the tools of measurement with the observed phenomena cannot be neglected in quantum theory, and "this point has not escaped attention in the work on the development of the quantum theory especially as regards problems of atomic constitution. Just recently, however, [this point] has been stressed […] by Heisenberg in connection with […] the symbolic method[s] developed in the last years and which have proved themselves [so] wonderfully suited for the elucidation of atomic problems." Almost the same words we find in the stenographic report of 9/16 where, after customary greetings and thanks to the participants and a few words in memory of Volta, Bohr acknowledges a "fundamental [limitation] in the classical idea[s] as regards the application [to] the atomic phenomen[a]." Here in the report he stresses the influence of the tools of measurement on the observed phenomena and notices: "This point [has not] escaped attention in the course of the development of the [quantum] theory especially as regards the problem of atomic constitution. [...] This fundamental point, however, has been discussed in a recent paper by H[eisenberg on] the commensurate [*sic*] development in the last years…"

In the manuscript of 9/13, Bohr now looks at the limitation of classical concepts from a "different point of view." Can phenomena be observed without being disturbed? Even if Rayleigh, Thomson, and Rutherford were able to obtain fundamental results on atoms and their constitution, the paradoxes of quantum theory "strikingly disclosed" the limitations of the classical ideas underling their works. The same concepts are recorded, though more synthetically, in the report of 9/16.

The second part of the manuscript of 9/13 (pp. 3-6) contains the core of Bohr's account. With reference to radiation processes, he recalls "Einstein's idea of individual light quanta, carrying energy and momenta expressed by the well-known quantum relations $E=h\nu$ and $P=h\sigma$." Energy and momentum go back to the idea of material particles, while frequency and wave-number imply the wave idea: "Indeed the wave and corpuscular ideas are able only to account for complementary sides of the phenomena." Note that Bohr just uses the adjective 'complementary,' not the noun 'complementarity' (see below, § 9). Also here, the stenographic report of 9/16 is similar in content, though more synthetic.

Next, in the manuscript of 9/13 Bohr switches from radiation processes to material particles, having an "individual character" and also behaving like waves. Here the similarities between the manuscript and the report become rather impressive (see the Appendix). Bohr makes reference to the recent experiments by Davisson and Germer[25] and invokes wave groups for describing particles, coming to the conclusion that "according to the quantum theory the possibility of a space time coordination is complementary to the possibility of a causal description." Similarly, though not mentioning causal description, in the report of 9/16: "We have a complementa[ry] connection between the coordination in those beams and the possibility of defining energy and momentum."

In the third part of the manuscript of 9/13 (pp. 6-7), Bohr goes through "a few simple examples," centred on a beam (of light or of electrons) passing through a hole. He finds $\Delta p\, \Delta l \sim h$, through elementary passages and with the help of a simple diagram. The same diagram, and elementary calculations now leading to the relation $\Delta p = h/\Delta l$, are recorded in the stenographic report of 9/16. This topic concludes the developed part of the manuscript of 9/13. The substance of Bohr's manuscript *Fundamental problems of the quantum theory* corresponds to the first two sections of *The quantum postulate and the recent development of atomic theory* on the Proceedings of the Volta Conference, and so does the stenographic report of 9/16.

A bare list of topics follows in the eighth page of the manuscript of 9/13: "Suggestion of statistical character of conservation. Disproved. Solution by wave theory..." and so on. This is the only point that Bohr briefly expands before his audience on 9/16: "[Discussions in the recent years have led] to the supposition that we have to be content with a statistical foundation for the [conservation] principles. As regards the [transmission] of energy and momentum this suggestion has proved to be wrong." Also the final passage of the stenographic report of 9/16 was missing in the manuscript of 9/13: "All these things are almost commonplaces. If you will have patience I shall try to show that this same state as over we meet everywhere." It seems that Bohr aimed to provide some more examples and applications of his views.[26] However, the report ends with Bohr's claim for some patience more.

---

[25] Davisson, & Germer (1927).

[26] Note that here again Bohr defines "commonplaces" his views (see note no. 13).

His time perhaps was over. Interestingly enough, reading aloud – like before an audience – the three developed parts of Bohr's manuscript of September 13 will take about twenty minutes.

The conclusions of our collation are straightforward: in delivering his speech in Como on September 16 Bohr stuck to the manuscript of September 13, without entering into further details before his audience.

## 6. In retrospect

Concise discussions by Born, Kramers, Heisenberg, Fermi, again Heisenberg, and Pauli[27] follow Bohr's paper on the Proceedings of the Volta Conference (which we recall were published in 1928). Along the above lines, it becomes worth noting that these discussions deal with only the first two sections of Bohr's paper, and show no relation with the content of the following three sections.

Our conclusions, that Bohr had only sketched his ideas on complementary features of quantum theory at the Volta Conference, make it obvious that "the reception of Bohr's presentation of his new ideas by the distinguished audience was remarkably cool."[28] From the point of view of Kalckar, who considers Bohr's speech ranging over the first four sections of the paper he would publish in 1928 if not the whole five, such cool reception may sound a bit curious; similarly curious it possibly sounds to M. Jammer that "in the discussion, in which Born, Kramers, Heisenberg, Fermi and Pauli participated, no objections were voiced".[29] Instead, we are now in a position to extend to Bohr's talk in Como a comment that Klein originally conceived for Bohr's manuscript of September 13: it "was so short that nobody could have understood it really."[30]

These circumstances recall to us Niels' brother Harald Bohr, who once "was asked why he was one of the greatest mathematical lecturer in the world while Niels was such an unsuccessful public speaker. He answered, 'Simply because at each place in my lecture I speak only about those things which I have explained before, but Niels usually talks about things he means to explain later'."[31] It should not be by chance that, on September 17, Lorentz regretted that discussions had been short of time in Como: "Then we have discussed the theory of 'quanta': unfortunately, we confined this problem to only the last session. [...] Yesterday, with the marvellous clarity and simplicity so well distinguishing himself, has Mr. Bohr given us a new ingenious explanation of that new mechanics of 'quanta'. We regret that we have not had time to undertake a thorough discussion, but we have all

---

[27] *Atti del Congresso internazionale dei fisici* (1928, vol. II, pp. 589-98).

[28] Kalckar (1985, p. 29).

[29] Jammer, M. (1966, p. 354).

[30] Klein, O. (1963, p. 11).

[31] Richard Courant, quoted in Pais (1991, p. 45).

the same heard and realised things we shall at ease reflect about once we are back to our sites."[32]

However, proving that Bohr's speech in Como in September was very concise, and expounded only embryonic ideas about complementary aspects of quantum theory, should not be an end in itself. It demands that the circumstances in which Bohr improved the presentation of his ideas should be clarified.

**7. From Como to Brussels**

The embryonic stage of Bohr's presentation in Como raises the question of appreciating the path toward his final presentation of complementary sides of nature, in order to enlighten if, in particular, interaction with various colleagues of his had any meaningful role in that path. Here we shall again consider letters, another manuscript of Bohr's – dated October 12-13 – and the stenographic reports of the discussions at the Solvay Meeting,[33] together with the first comprehensive paper that Bohr finally published in 1928.[34]

Kalckar himself recalls that instead of leaving for Copenhagen, "after the Volta Meeting in Como, Bohr spent a week together with Pauli at Lake Como[35] in order to prepare the publication of an extended version of his lecture." Conversations with Pauli had been an acknowledged stimulus for Bohr; neither was Pauli new to assisting Bohr in formulating some of his papers.[36] However, from Kalckar's point of view, a prolonged interaction after the Conference would just imply that Pauli gave help only adjusting an already comprehensive, developed and almost exhaustive presentation of complementarity, relating to at least four of the five sections of Bohr's forthcoming paper. It is straightforward that, given the real, meagre content of Bohr's talk in Como, conversation with Pauli might instead inspire and help Bohr to elaborate the third and fourth paragraph; that is, conversations with Pauli – who would be in fact the first to define wave-particle duality in a publication[37] – might indeed give a fundamental contribution to the appearance of Bohr's final presentation of complementary sides of nature. Besides Pauli, most probably Bohr met also with Darwin, who took part in the Volta Conference on behalf of the Royal Society of Edinburgh. From another letter from Edinburgh on October 6, we can infer that Darwin — author of a paper where, as far as I know, Bohr's idea of complementary aspects of atomic theory is published for the first

---

[32] *Atti del Congresso internazionale dei fisici* (1928, vol. II, p. 625).

[33] *Notes from Solvay Meeting* (1927).

[34] Bohr (1928a).

[35] Actually, after September 16 the Conference moved to Pavia and then to Rome, where the conclusive ceremony took place on September 19.

[36] See for example Bohr's letter to Heisenberg of 18/4/1925, cited in Mehra & Rechenberg (2000, p. 163; see

time[38] — did not leave for home before Saturday, September 24,[39] so that he could well have, in a spirit of reciprocal enrichment, five days for private discussions with Bohr after the end of the Conference. That is, also conversations with Darwin might possibly substantiate Bohr's ideas on complementary sides of nature before the Solvay Meeting.[40]

On October 11 Bohr sent a new manuscript, in German, to *Naturwissenschaften* for publication and to Pauli for critical remarks (contextually, Bohr asked the editors of the journal to send Pauli a copy of the proofs).[41] The manuscript in German is lost, but we can all the same read Pauli's reply to Bohr in a letter from Hamburg of October 17.[42] Bohr also prepared an English version of the manuscript, dated October 12-13. Then, on October 16 he wrote to Darwin: "I enclose a note which I sent to *Nature* just a few days ago. I am not very satisfied with it [...] and I had to postpone the discussions of examples to the Como Lecture."[43] The English manuscript of October 16 has the same title *The quantum postulate and the recent development of atomic theory* as Bohr's paper to come.[44] It is much more elaborated than the manuscript titled *Fundamental problems of the quantum theory* of September 13, so that we cannot share Mehra and Rechenberg's opinion that the "English manuscript [of October 16] and the published paper [...] do provide a fair impression of what Bohr actually presented at Como"[45] — on the contrary, the meagre content of Bohr's speech in Como gives the measure of the progress involved with the manuscript sent to Darwin (and to Pauli).

---

also p. 107).

[37] Pauli (1933).

[38] Darwin (1927). Darwin's paper, received by the Journal on 10/25/1927, opens acknowledging that "The author has had the advantage of many conversations with Prof. N. Bohr on the subject" (p. 258). In turn, Bohr mentions Darwin's forthcoming paper at the Solvay Meeting, in answer to a question by Lorentz arising from the Stern-Gerlac experiment. *Notes from Solvay Meeting* (1927), General discussion, typewritten transcription of the session of Thursday 10/27, p. 2, n. 20.

[39] "My dear Bohr, [... my wife and I] had a slow journey home travelling only by day and so taking three days. Then I had a week in Cambridge and London [...]. I only arrived on Monday [October 3]." (Darwin to Bohr, from Edinburgh, on 10/6/1927; *NBSC*, microfilm n. 9). Accordingly, Darwin reached Cambridge one week before Edinburgh, on Monday, September 26, having travelled for three days, presumably since Saturday, September 24. We note in passing that in this same letter Darwin mentions a paper of his, devoted to wave mechanics and on the way to being finished (Darwin 1927).

[40] To my knowledge, this possibility is not acknowledged in the literature, which ignores that Bohr's presentation was still embryonic in Como.

[41] Kalckar (1985, p. 30).

[42] Kalckar (1985, pp. 32-35).

[43] Letter from Bohr to Darwin of October 16, 1927. *Archive for the History of Quantum Physics*. A microfilmed copy of the *AHQP* is owned by the Accademia nazionale delle scienze detta dei XL in Rome (the letter is on microfilm no. 36).

[44] Kalckar (1985, pp. 91-98).

[45] Mehra & Rechenberg (2000, p. 196).

Instead of being limited to only the first two sections of the paper that would be published on the Proceedings of the Volta Conference, the October manuscript covers many of the topics included in the third and fourth sections (even touching upon some topics of the last, fifth section).

Besides worthy similarities between the 1928 paper and the October manuscript, differences are also very significant. In the Proceedings of the Volta Conference Bohr analyses in detail some experimental facts and thoroughly accounts for the generalities of the wave description, even though he had not done the same in his manuscript of October 12-13.

At the end of the second section of the 1928 paper, Bohr discusses the impossibility in principle of determining the velocity of a particle by determining its positions at two given moments.[46] Kalckar does not notice that this topic was absent in the October manuscript. It is worth noting now that such impossibility in principle was specifically one of the topics discussed in Brussels, as testified in a letter of Ehrenfest to Goudsmit, Uhlenbeck and Dieke.[47]

In that same letter Ehrenfest writes that Bohr, basing on the energy conservation principle, extends the uncertainty relations from light to material particles. The stenographic report of the Solvay Meeting gives evidence of Ehrenfest's account. Now, this very extension to material particles was initially ignored in Bohr's manuscript of October 1927, whereas it is included in Bohr's paper of 1928.[48]

As early as 1925 Bohr had investigated the scattering of fast alpha-particles by atoms. In fact, he discussed it again with Lorentz at the Solvay Conference. Here again, the scattering of fast alpha-particles was not included in the October manuscript, but Bohr includes it in his 1928 paper.[49]

Further, the Stern-Gerlach experiment:[50] Bohr did not mention it in his manuscript of October 12-13, but Ehrenfest urged him and Lorentz to expound on it at the Solvay Meeting. Eventually Bohr included the Stern-Gerlach effect in his 1928 paper.[51]

The stenographic reports record Bohr's wide exposition in the General discussions of October 27

---

[46] Bohr (1928a, pp. 574-575).

[47] Letter from Ehrenfest to Goudsmit, Uhlenbeck and Dieke, dated Leiden, November 3, 1927. Kalckar (1985, pp. 37-41).

[48] *Ibid.*, in Kalckar (1985, p. 39). *Notes from Solvay Meeting* (1927), General discussion transcribed by Verschaffelt of Thursday 10/27, p. 4. Bohr (1928a pp. 571-572).

[49] Bohr (1925, p. 848). *Notes from Solvay Meeting* (1927), General discussion of Thursday 10/27, p. 2, nn. 24-27, and session of Friday 10/28 transcribed by Verschaffelt p. 1. Bohr (1928a, p. 583).

[50] Note that on February 1927 Pauli discussed its importance with Heisenberg, who in turn included it in his paper on the uncertainty relations. Mehra & Rechenberg (2000, p. 157). Heisenberg (1927).

[51] *Notes from Solvay Meeting* (1927), General discussion, transcribed by Klein, O., p. K2 bis; transcription by Verschaffelt, pp. 4-5; typewritten transcription of the session of Thursday 10/27, p. 2, n. 19. Bohr (1928a, p. 584).

and 28,[52] to which also Lorentz, Brillouin, Kramers, Fowler, Einstein, Born, Ehrenfest, Pauli, Schrödinger, Debye, Dirac, Bragg, and Heisenberg took part with their remarks.[53] Bohr tackled the gamma-ray microscope and the momentum measurement by Doppler effect, which were not mentioned at all on his October 12-13 manuscript.[54] These two topics eventually take about three pages of the 1928 paper on the Proceedings of the Volta Conference.[55]

Bohr himself, on earlier occasions, had already tackled some of the topics that he expounded or discussed at the Solvay Meeting and then included in his paper on the Proceedings of the Volta Conference, thus improving his manuscript of October 12-13. The already investigated scattering of fast alpha-particles for example; but also the indeterminacy in the wavelength and in the parallelism of the waves or the Doppler effect, which Bohr had once introduced with Einstein, as well as the gamma-ray microscope discussed with Heisenberg.[56] It might be partly as a consequence of his individual efforts to work out a more detailed exposition than his speech in Como, partly as a consequence of the discussions in Brussels, that Bohr resolved how effective these subjects might be in elucidating his idea of complementary features of the description of experience.

## 8. Physicists in dialogue

M.J. Klein, in his *The first phase of the Bohr-Einstein dialogue*, investigates the inauguration of a controversy between Einstein and Bohr at the time when Bohr, Kramers and Slater invoked virtual radiation in order to preserve the wave theory of radiation.[57] Later, Bohr's letter to Einstein of April 13, 1927 is acknowledged as having "opened up a new Bohr-Einstein dialogue that would reach its first climax at the Solvay Conference in the following October".[58] Yet, we hardly find evidence of any Bohr-Einstein dialogue in the Proceedings of the Solvay Conference published in 1928, where very few discussion remarks by Einstein – who had not been in Como in September – and Bohr are recorded. There are various reasons for this lack of evidence. One is quite general: the Proceedings are not an exhaustive, literal report of the discussions among the participants, as is disclosed by the stenographic reports; rather they are a rationalised account. Further, almost all of Bohr's remarks at the Conference are included in his paper, instead of being recorded separately. Moreover, if "very little of what [the participants] said came through in the official published discussions",[59] it was also because discussions — particularly those between Einstein and Bohr — were often referred to 'unofficial' sessions:

> Every night at 1 a.m. Bohr came into my room just to say ONE SINGLE WORD to me, until three a.m. It was delightful for me to be present during the conversations between Bohr and Einstein. Like a game of chess. Einstein all the time with examples. In a certain sense a sort of Perpetuum Mobile of the second kind to break the UNCERTAINTY RELATION. Bohr from out of philosophical smoke clouds constantly searching for the tools to crush one example after the other.[60]

---

[52] *Notes from Solvay Meeting* (1927), General discussion transcribed by Verschaffelt, sessions of Thursday 10/27, p. 3, and of Friday 10/28, p. 3.

[53] *Notes from Solvay Meeting* (1927), General discussion, typewritten, sessions of Thursday 10/27 and of Friday 10/28.

[54] We however recall that Bohr had already mentioned the Doppler effect in his letter to Einstein of April 13.

[55] Bohr (1928a, pp. 572-574).

[56] Letter to Einstein of April 13, 1927 (Kalckar 1985, pp. 21-24).

[57] Klein, M.J. (1970). For further accounts of the role of dialogues in the development of quantum physics, see for example Klein (1970), Hendry (1984), Beller (1999), and also Heisenberg (1971, p. xvii).

[58] Mehra & Rechenberg (2000, p. 188).

Any attempt to investigate the Einstein-Bohr discussions in Brussels comes out to be crucial and problematic at the same time. Only seldom are Einstein's remarks recorded in the stenographic report of the Solvay Meeting, if not merely mentioned with a laconic "Einstein having exposed his *Standpunkt*, Bohr says...".[61] We perceive Einstein's sceptical vein: "Einstein asks Bohr if he may express his ideas with ordinary words, avoiding mathematical formulas difficult to interpret. If one might represent facts in a less childish way, maybe the question becomes clearer."[62] Yet, to Ehrenfest's delight, he did not renounce to deepen the question with Bohr privately, thus meeting the latter's early wishes: "How nice it would be once again to talk to you face to face about all these things."[63]

Kalckar attempts to portray fragments of the discussions that involved Einstein, by alternating excerpts from the official Proceedings and from the stenographic reports of the Solvay Conference.[64] There, we can read about Einstein's criticism of the quantum theoretical description of an electron's passage through a slit and about Bohr's answer[65] — the former recorded among the discussion remarks in the Proceedings published in 1928, the latter in the stenographic report written during the Meeting. We meet however a serious drawback: a detailed analysis of the discussions in Brussels cannot be firmly based on the official Proceedings, since their conformity to the stenographic reports is not obvious. For example, even if the Proceedings seem to suggest that "Einstein remained silent after the presentations of Compton and Bragg",[66] in the stenographic report one can find two brief remarks of Einstein's in the discussion following Compton's speech (in addition to the biting comment Einstein addressed to Bohr, we already mentioned above).[67] Emphasis is also different: in the Proceedings we read of Heisenberg stating that "I do not agree with Dirac when he says that in the experiment described nature makes a choice".[68] In the stenographic report it is put differently: Bohr recalls with Dirac that "Nature makes a choice," Heisenberg promptly replies "If ... then nature would never make choice," and Bohr remarks "Choice of making it observable." In conclusion, Dirac "doesn't agree with Heisenberg quite" (not vice-versa).[69] On the whole, a somewhat confusing picture: "No answer from Bohr to Einstein's analysis of the electron's passage through a slit or screen was recorded", contrarily to what we noted above.[70]

---

[59] Mehra & Rechenberg (2000, p. 246).

[60] Letter from Eherenfest to Goudsmit, Uhlenbeck and Dieke, *cited*. Kalckar (1985, p. 38).

[61] *Notes from Solvay Meeting* (1927), transcription by Verschaffelt of the Continuation of General discussion on Friday 10/28, p. 3.

[62] *Notes from Solvay Meeting* (1927), discussion following Compton's speech, p. 2. This Einstein remark is cancelled in the report, but remains well readable.

[63] Letter of Bohr to Einstein of April 13, 1927 (Kalckar 1985, pp. 23-24).

[64] Kalckar (1985, pp. 99-106).

[65] Kalckar (1985, p. 103).

A detailed analysis of the stenographic reports of the Fifth Solvay Meeting lies beyond the scope of the present paper. Our aim is merely to review the path toward Bohr's presentation of complementary sides of nature in his 1928 papers and the possible role of the conversations he had with his colleagues. Our claim now is that in some cases a role of these dialogues is more plain — as seen in § 7 —, but portraying Einstein's possible involvement is much more problematic. Furthermore, we also claim that the official Proceedings should preferably not be kept separated from the stenographic reports, for a thorough investigation of the early debates on the complementary features of quantum theory.

In her *Quantum dialogue*, Mara Beller deals with what she calls *The dialogical birth of Bohr's complementarity*.[71] She claims that the efforts of scientists are fundamentally linked to one another. By analysing Bohr's 1928 papers and his colleagues' works — an example of what she calls "the intricate flux of dialogues among quantum physicists" —, she comes to the conclusion that Bohr's 1928 published papers should not be considered "the unfolding of a single argumentative structure, but as the juxtaposition of several simultaneously coexisting arguments, addressed to different quantum theorists about different issues".[72]

Our attempts to elucidate the dialogical framework in which Bohr conceived and presented his ideas about complementary features of quantum physics may appear to fully match Beller's arguing. However, even if we ourselves are dealing with this dialogical framework, we are tracing it in a perspective that is substantially different from Beller's.

---

[66] Mehra & Rechenberg (2000, p. 242).

[67] *Notes from Solvay Meeting* (1927), discussion following Compton's speech, p. 4. See also note no. 62.

[68] *Atti del Congresso internazionale dei fisici* (1928, p. 264). Reproduced in Kalckar (1985, p. 105).

[69] *Notes from Solvay Meeting* (1927), General discussion, typewritten transcription of the session of Friday 10/28, p. 5, nn. 51-55.

[70] Mehra & Rechenberg (2000, p. 250). But see note no. 65.

[71] Beller (1999, pp. 117-144).

Beller uses the word 'dialogue' in a figurative sense. Bohr's papers are imagined to be part of a 'dialogue', in which they substantially represent his public response to the work of his colleagues. But in his 'dialogues' with Schrödinger, Campbell or Pauli, Bohr writes his papers addressing them to the whole community of physicists of which the single Schrödinger, Campbell or Pauli are members. In turn, his peers reply individually and not collectively, to the whole community and not to the single Bohr. In other words, if we stick to the literal sense of the term dialogue, i.e. a conversation — even in public — between two or more individuals, then Beller's imaginative 'dialogues' lack reciprocity.

The different senses — figurative and literal — in which dialogues are conceived make Beller's approach and ours definitely diverge from the one another. Our claims are that Pauli and possibly Darwin personally joined Bohr in discussing how the presentation of his views might be substantially improved, and that in his 1928 papers we can point at the signs of fruitful discussions he had reciprocally had with his peers in Brussels in October 1927 — no matter if early at breakfast, walking from the hotel to the conference building in the morning, in the conference hall during the day, or in Ehrenfest's room at night.

Beller's imaginative 'dialogues' are implicit: she aims at "uncovering and describing the underlying network of implicit scientific dialogues in the Como lecture", a network that "has to be uncovered rather than simply pointed at".[73] Instead, we dealt with Bohr's explicitly mentioned or recorded conversations in Como and in Brussels.

## 9. Refining upon the style

I would like now to add just a few remarks about the well-known circumstance that, *after* the Solvay Conference, various colleagues continued to help Bohr to give "The final touch" to his work.[74] In 1928 Bohr wrote another detailed piece of work on the complementary aspects of atomic physics, further expanding the previous paper on the Proceedings of the Volta Conference – which we stress again was published in 1928 too. He published it in four versions: in English for *Nature*, in French for the Proceedings of the Solvay Conference, in German for *Naturwissenschaften*, and, in 1929, in Danish.[75] Dirac's involvement, and most relevantly Pauli's continued help with the proofs are documented by a few letters ranging until March 1928.[76] But also Fowler and Hartree were involved with translation and proofs correction. Bohr's letters with them are symptomatic of some trouble with the manuscript. Bohr, who in October had already confided to Darwin he was not satisfied with his own manuscript, on December 27 updates Fowler "about the fate of my article with which you helped me so kindly in Cambridge. The re-modelling of my article has taken more time than I expected. I have put a lot of work into it and have finally rewritten the whole manuscript such as it was suggested to me by the editor of *Nature*. [...] I have asked him to send you a proof."[77] On these same lines, three months later Bohr wrote to Hartree: "I was very thankful for all your kind help with the translation of my article when I was last in Cambridge. Since that time I have put a great deal of work into it trying to improve the representation of my views, and only a week ago I have returned my final proof to *Nature*."[78] In his reply to Bohr, in January 1928, Fowler comments on Bohr's neologism: "Your new word Complementarity is very nice [...]. But I do not believe it exists. Complementary nature is all I can suggest for it."[79] It is the first time that the use of "the new word Complementarity" by Bohr is (indirectly) documented. In all available manuscripts of Bohr's dated 1927, he had always used the adjective 'complementary' in place of the abstract noun 'complementarity:' complementary aspect of experience, complementary sides, complementary features, complementary nature, complementary ideas.[80]

---

[72] Beller (1999, pp. 2, 8).

[73] Beller (1992, pp. 148, 151). See also Beller (1999, p. 120).

[74] Kalckar (1985, pp. 41-53). See also Mehra & Rechenberg (2000, pp. 256-60).

[75] Bohr (1928b,c,d; 1929). Pais mistakenly states that the shorter paper on the Proceedings of the Volta

## 10. Conclusions

Concerning Bohr's Como lecture, Beller nicely perceives that it is "populated with invisible interlocutors".[81] Strange as it may appear, referring to the "Como lecture" in the literature may

---

Conference, and not the expanded version on *Nature*, was translated in French for the Proceedings of the Solvay Conference (Pais 1991, note on p. 318). In any case, it should be borne in mind here that when Bohr and his colleagues make generic reference to the proofs of an English manuscript, we are not in a position to distinguish between the two versions, unless *Nature* or Como are explicitly mentioned by the correspondents (Bohr 1928a,b).

[76] Kalckar (1985, pp. 41-46).

[77] Letter of Bohr to Fowler of 12/27/1927. *NBSC*, microfilm no. 9.

[78] Letter of Bohr to Hartree of 3/27/1928 (underlining in the original). See also the letter of Hartree to Bohr of 4/3/1928. *NBSC*, microfilm no. 9.

[79] Letter of Fowler to Bohr of 1/24/1928. *NBSC*, microfilm no. 9. Text underlined in the original.

[80] Kalckar (1985, pp. 61, 62, 69, 76, 91, 93, 94, 96). Kalckar transcribes "complementarity" once, but on the manuscript of September 13 it more closely resembles "complementary" than "complementarity." Kalckar (1985, pp. 78, 86). Also Pais mistakenly states that "the term 'complementarity' appears for the first time in a draft from 10 July 1927. In Bohr's correspondence it shows up in a letter to Pauli in August." Instead, on 7/10 Bohr writes "complementary aspects of experience," while on 8/13 he only writes to Pauli "complementary sides of nature" and "complementary sides of the question" (komplementære Sider hos Naturen, komplementære Sider af Sagen) (Pais 1991, p. 311; Kalckar 1985, pp. 61-62; Pauli 1979, p. 406).

[81] Beller (1999, p. 117).

generate confusion in the reader. It is generally meant to be the paper Bohr published on the Proceedings of the Volta Conference, but its connection with the real content of Bohr's talk in Como had never been seriously questioned. Serious misleading arose: a substantial correspondence of the Como (published) lecture with Bohr's Como speech (at the conference) is tacitly assumed, irrespective of the meagre content of a manuscript Bohr wrote in Como on September 13 and of the stenographic report of Bohr's speech which "seems incomprehensible".[82] However, a clear correspondence between Bohr's manuscript of September 13 and the stenographic report of his speech in Como can be established, so disproving that Bohr thoroughly accounted for complementarity already at the Volta Conference.[83] Bohr's speech in Como stuck to the manuscript; that is: his speech "was so short that nobody could have understood it really."[84]

The manuscript of September 13 thus becomes much more significant than has hitherto been realised. Another fundamental manuscript, of October 12-13, records the provisional stage of Bohr's ideas previous to the Solvay Meeting of October 24-29. It had preliminarily been discussed by Bohr with Pauli and most probably with Darwin, before they all left Italy after participating in the Volta Conference. If we conform to the old assumption that Bohr gave a thorough lecture already at the Volta Conference, we would hardly attach fundamental significance to conversations in the intervening period between the Como and Brussels conferences: Kalckar actually acknowledges Pauli's aid in the lead-up to the Solvay Conference, but he cannot adequately highlight that Pauli's (and Darwin's) aid might possibly concern the formulation of the third and fourth paragraphs of Bohr's paper from the beginning.[85]

Concerning Bohr's so called Como lecture (i.e. his paper in the Proceedings of the Volta Conference), we should mind not only that it differs substantially from his speech in Como, but also that it was published in 1928, after he had taken part in the Solvay Conference. This timing implies that, besides his individual efforts to complete the presentation to be delivered before his colleagues, also the conversations Bohr had with his peers in Brussels might have some influence on his Como lecture, in a similar way as those with Pauli and Darwin possibly had. As a matter of fact many elucidatory issues, ignored in his October manuscript, after the Solvay Conference were judged substantial enough to be included in his Como lecture, for example many real (e.g. the Stern-Gerlach) and thought (e.g. the gamma-ray microscope) experiments. All the more so, the final version of his work (i.e. the version in the Proceedings of the Solvay Conference and the *Nature* article) goes very far beyond Bohr's speech in Como in September — even if the well known interactions with his colleagues in the follow up of the Solvay Conference in October possibly did not lead to as substantial addictions as those in dialogue with Pauli, Darwin, and later from Brussels.

---

[82] See note no. 24.

[83] Bohr himself might put on the wrong track, as far as he writes in a note on *Nature*: "The content of this paper is essentially the same as that of a lecture on the present state of quantum theory delivered on Sept. 16, 1927, at the Volta celebration in Como" (Bohr 1928,b, p. 580 note no. 1). However, "essentially" should be meant as "basically" here: the September lecture was a starting point to which more has been added. Similarly, at the end of his October manuscript Bohr had written: "A more detailed elaboration of this point of view in its application to a number of simple examples was recently given by the author in a lecture at the Volta congress" (see Kalckar (1985, p. 98)).

[84] See note no. 29.

[85] See Kalckar (1985, pp. 29-35).

All that given, we cannot anymore subscribe Kalckar's claim that, in particular, "the discussion with Einstein at the Solvay Meeting did not have any major impact on the final elaboration of the text in the form in which it appears in the transactions of the conference"[86] (and, if not Einstein, one might even have wondered who else could basically affect Bohr's presentation of complementarity). In fact, as Mehra and Recheneberg warn, "one should not play down too much the role of outside influences on Bohr's conception of complementarity. In any case, the detailed investigation of the events that led Bohr from the middle of 1925 onward to the final formulation of complementarity in late 1927 or early 1928 reveals a very complex interrelationship of ideas and results".[87] In the present paper we tried to throw some new light on this complex web of interrelationship, on the basis of documental evidence. We hope it may be a fruitful line of research for further deepening the genesis of Bohr's work on complementarity but also more in general, for investigating other works of Bohr's or of other physicists'.

**Acknowledgement**

I am indebted with the *Niels Bohr Archive* for letting me consult Bohr's letters with the Committee of the Volta Conference and the transcription of the discussions at the 1927 Solvay Meeting, and to Dr. Felicity Pors for her helpful availability. I am also grateful to the *Accademia Nazionale delle Scienze detta dei XL*, for letting me consult the microfilmed copy of Bohr's letters and of the stenographic transcription of Bohr's speech in Como, and to Dr. Antonella Grandolini for her most kind help. I am also grateful to Prof. Fabio Sebastiani for his continuous encouragement.

---

[86] Kalckar (1985, p. 32). On comparing the Como lecture with the expanded version on *Nature*, Kalckar adds: "There are no additions that warrant the conclusion that they are direct results of the discussions with Einstein at the Solvay Meeting" (ibid., p. 44). However, the Como lecture is not a good term of comparison for the expanded version, as far as discussions in Brussels are to be found already in the former.

[87] Mehra & Rechenberg (2000, p. 164).

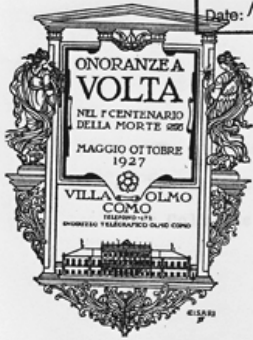

**Caption** Letter of the Committee for the Celebrations in honour of Volta to Bohr, of 7/4/1927 – Courtesy Niels Bohr Archive.

# APPENDIX

On the left, the text of Bohr's manuscript of September 13. On the right, the stenographic report of September 16, with some corrections suggested by the context (for example, *Compton theory* and *supposition* recorded by the stenographer have been corrected with *quantum theory* and *superposition*).

| page | **Bohr's Manuscript: September 13** | **Stenographic report of Bohr's speech: September 16** | page |
|---|---|---|---|
| 1 | *Characteristic of the quantum theory is the acknowledgement of a fundamental limitation in our classical physical ideas when applied to atomic phenomena.* | *Characteristic for the quantum mechanic[s] is the [acknowledgement] of the fundamental [limitation] in the classical idea[s] as regards the application [to] the atomic phenomen[a].* | 1 |
| | *Indeed our usual space time coordination rests entirely on the idea of tools of measurement [the] interaction of which with the phenomena to be observed may be neglected.* | *[Indeed our] usual description rests essentially on the idea that we possess the possibility of observing with almost [neglecting] the influence o[n] the phenomen[a] that we are studying in our observations.* | |
| 2 | *This point has not escaped attention in the work on the development of the quantum theory especially as regards problems of atomic constitution. Just recently, however, it has been stressed in a very interesting and suggestive way by Heisenberg in connection with [...] the symbolic method[s] developed in the last years and which have proved themselves [so] wonderfully suited for the elucidation of atomic problems.* | *This point [has not] escaped attention in the course of the development of the [quantum] theory especially as regards the problem of atomic constitution. [...] This fundamental point, however, has been discussed in a recent paper by H. [on] the [symbolic methods?] developed in the last years [and which] will permit [...] analysing the atomic phenomen[a].* | 2 |
| | *Now the modern development of science depends on the applicability of these methods also [to] the atomic phenomena.* | *With great success it has been possible to apply the theory of ... to atomic phenomen[a]* | |
| 3 | *Notwithstanding the success of the wave theory it has not been possible to account for interchanges of momentum and energy by radiation processes except by Einstein's idea of individual light quanta, carrying energy and momenta expressed by the well-known quantum relations $E=h\nu$ and $P=h\sigma$. [...] Indeed the wave and corpuscular ideas are able only to account for complementary sides of the phenomena.* | *We must make use of the idea of [light] atoms. Einstein said that energy is equal to $E=h\nu$ [and the momentum] $P=h[\sigma]$.* | 2-3 |
| 3-4 | *On the other hand the formulas [do?] not only express the [individual?] character of the elementary radiation processes [but in this way?] the definition of energy and momentum may be carried back to the idea of material particles [...] Notwithstanding the very direct way in which the individual character of the electrons is brought out by the evidence [...], the discovery of Davisson* | *On the other hand the energy momentum goes back to the idea of materia[l particles]. We have of course very convincing evidence as regards the individuality of electrons. We cannot account for a reflection of electrons discovered by [Davisson and Germer, if we do not make use] of the wave [superposition] principle. The experiments are most wonderfully in agreement with the ideas of* | 3 |
| | *and Germer of the selective reflection of electrons from metal crystals prove[s] the necessity of applying a wave theoretical superposition principle [...]. As well-known the experiments are in complete accordance with the ideas of de Broglie.* | *[de Broglie].* | |
| 5 | *The term within the bracket is nothing else, than the negative value of the scalar product of the space time vector [...] and the Impulse Energy vector [...]. As emphasized by de Broglie the abstract character of the phase-wave is [already?] indicated by the fact that its velocity of propagation $v^x$ is always larger than the velocity of light c [...] and the only way of observing an elementary wave is by interference.* | *This expression is nothing else but the product of the [space-time vector and the impulse-energy vector]. Now as [stressed] by [de Broglie] this wave is an abstraction ... v = [...] These waves are obtainable by interference of other waves.* | 4 |
| 6 | *a limitation of the group in extension in space and time is [...] conjugated to a limitation in accuracy with which energy and momentum can be defined. Indeed we may say that according to the quantum theory the possibility of a space time coordination is complementary to the possibility of a causal description.* | *We have a complementa[ry] connection between the coordination in those beams and the possibility of defining energy and momentum* | 5 |
| | *if we open the [hole?] a time t the frequency is only defined by $\Delta v = 1/t$ and the energy of the light quantum and the electron is therefore only known by an accuracy given by $\Delta t\, \Delta E = h$.* | *if we open this window only a short time we cannot give the frequency of the waves. We cannot know how large the energy of the light quantum is.* | |
| | *Let us in order to know the energy not care about the time and let the hole left open. [...] If the diameter of the hole is l the outpassing wave will be defracted over an angle of magnitude $\alpha = \lambda/l$. [...] The momentum is $h/\lambda = p$. The component parallel to the hole however will be undetermined to the amount $\Delta p = \alpha p$ also [thus] $\Delta p\, \Delta l \sim h$.* | *if we kept the window open then we would know what the frequency of the wave is. If the window of a[perture l...] what we will do when the [wave] will be [diffracted] and we get uncertainty [by] the calculation $\Delta p = h\tau\alpha = h/l$.* | |
| 7 | *Of course these illustrations give nothing new [... They] show however clearly how impossible it is in experimental arrangements to go beyond the limitations discussed.* | *This consideration means nothing [new]. We are working in the limit of defining such quantities [as] energy and momentum* | 6 |
| 8 | *Suggestion of statistical character of conservation. Disproved. Solution by wave theory...* | *[Discussions in the recent years have led] to the supposition that we have to be content with a statistical foundation for the [conservation] principles. As regards the [transmission] of energy and momentum this suggestion has proved to be wrong* | |